\documentclass[twocolumn,aps,prl,floats,superscriptaddress,longbibliography]{revtex4-2}
\usepackage{epsfig}
\usepackage{graphicx}
\usepackage[figuresright]{rotating}
\usepackage{relsize}%
\usepackage[version=4]{mhchem}

\usepackage{amsfonts}
\usepackage{amsmath}
\usepackage{amssymb}
\usepackage{wasysym}

\usepackage{pifont}
\usepackage{mathtools}
\usepackage{bm}
\usepackage{cases}
\usepackage{braket}

\PassOptionsToPackage{normalem}{ulem}
\usepackage{ulem}
\usepackage[unicode=true,pdfusetitle,
 bookmarks=true,bookmarksnumbered=false,bookmarksopen=false,
 breaklinks=false,pdfborder={0 0 1},backref=false,colorlinks=false]
 {hyperref}
\hypersetup{colorlinks,linkcolor=blue,citecolor=blue,urlcolor=blue}

\makeatletter

\usepackage{xcolor}
\usepackage{pdfcolmk}
\usepackage{wasysym}

\usepackage{txfonts}

\usepackage{microtype}
 
\makeatother

\newcommand{\Chi}{\mathrm{X}}
\begin{document}

\title{Stacking-induced magnetic frustration and spiral spin liquid}
\author{Jianqiao Liu}
\thanks{These authors contributed equally.}
\affiliation{State Key Laboratory of Surface Physics and
Department of Physics, Fudan University, Shanghai 200433, China}

\author{Xu-Ping Yao}
\thanks{These authors contributed equally.}
\affiliation{Department of Physics and HKU-UCAS Joint Institute 
for Theoretical and Computational Physics at Hong Kong, 
The University of Hong Kong, Hong Kong, China}

\author{Gang Chen}
\email{gangchen@hku.hk}
\affiliation{Department of Physics and HKU-UCAS Joint Institute 
for Theoretical and Computational Physics at Hong Kong, 
The University of Hong Kong, Hong Kong, China}
\affiliation{The University of Hong Kong Shenzhen Institute of Research and Innovation, Shenzhen 518057, China}

\date{\today}

\begin{abstract}
Like the twisting control in magic-angle twisted bilayer graphene, the stacking control 
is another mechanical approach to manipulate the fundamental properties of solids,
especially the van der Waals materials. 
We explore the stacking-induced magnetic frustration and the spiral spin liquid 
on a multilayer triangular lattice antiferromagnet where the system is built from 
ABC stacking with competing intralayer and interlayers couplings. 	
By combining the nematic bond theory and the self-consistent Gaussian approximation, 
we establish the phase diagram for this ABC-stacked multilayer magnet. 
It is shown that, the system supports a wide regime of spiral spin liquid 
with multiple degenerate spiral lines in the reciprocal space, separating 
the low-temperature spiral order and the high-temperature featureless paramagnet. 
The transition to the spiral order from the spiral spin liquid regime is first order.
We further show that the spiral-spin-liquid behavior persists even 
with small perturbations such as further neighbor intralayer exchanges. 
The connection to the ABC-stacked magnets, 
the effects of Ising or planar spin anisotropy, and the outlook  
on the stacking-engineered quantum magnets are discussed. 	
\end{abstract}

\maketitle


Since the discovery of superconductivity~\cite{Cao2018B}, quantum anomalous Hall effect~\cite{doi:10.1126/science.aay5533} and other phenomena~\cite{Cao2018A,doi:10.1126/science.aav1910,doi:10.1126/science.aaw3780,Wong2020,Nuckolls2020,Choi2021,Saito2021,Hesp2021,Xie2021} 
in twisted bilayer graphene, twistronics has emerged as an important and popular field in the study of 
two-dimensional (2D) materials. The crystal twisting provides an important control knob to manipulate 
the electronic properties of quantum materials and also to induce exotic quantum 
phases of matter in the underlying electronic systems. Like the more popular twisting scheme,
the stacking control is another useful structural manipulation of the stacking orders 
of 2D materials through rotation and translation between the layers. 
The stacking procedure has been successfully used to manipulate the electronic 
and optical properties of layered van der Waals (vdW) materials~\cite{Bao2011,Lui2011,doi:10.1126/sciadv.aat0074,Jiang2014}, and the application to
the 2D magnetism has recently been explored~\cite{doi:10.1126/science.aav1937,Li2019,Song2019}. 
Modern fabrication techniques such as mechanical exfoliation~\cite{Huang2017,Gong2017,Fei2018,Deng2018,doi:10.1021/acs.nanolett.6b03052,Wang_2016} and 
molecular beam epitaxy~\cite{doi:10.1021/acs.nanolett.8b00683,Bonilla2018} make such a stacking control of magnetism feasible.
It was shown that, the interlayer coupling depends strongly on the stacking, 
allowing the manipulation of the magnetic properties of the stacked magnets~\cite{doi:10.1126/science.aav1937,Li2019,Song2019}.  
While existing works focus on the different magnetic orders resulting from the 
stacking, in this Letter we explore the possibility of stacking-induced 
magnetic frustration as well as liquid-like fluctuating regimes from frustration.

\begin{figure*}[t]
	\includegraphics[width=1.0\linewidth]{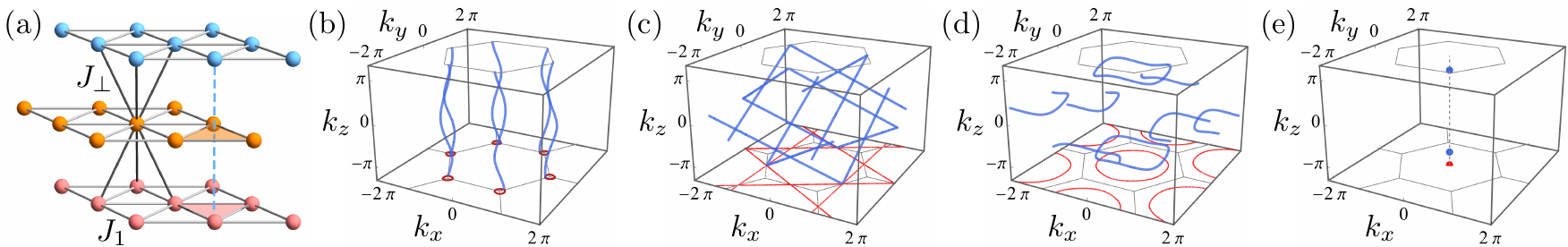}
	\caption{(a) The multilayer triangular lattice with the ABC stacking. 
	The dashed line along the $c$ direction indicates the projection of a site from the top layer to the centers of unequivalent triangles within the lower two layers. 
	The intralayer and interlayer interactions are denoted by $J_1$ and $J_{\perp}$, respectively. 
	The spiral manifolds (blue) and their projections (red) on the $k_x$-$k_y$ plane are presented for 
	(b) ${J_\perp/J_1=0.3}$, (c) ${1.0}$, (d) ${1.5}$, and (e) ${3.0}$.
	The BZ boundaries for a monolayer triangular lattice are plotted in gray.}%
	\label{fig:lattice_and_groundstates}
\end{figure*}

We start from the 2D magnet with the simplest frustrated structure, i.e., the triangular lattice, 
and stack the triangular layers along the $c$ direction to form a multilayer three-dimensional (3D) system. 
The stacking order was known to be crucial in determining the 
electronic states~\cite{PhysRevLett.104.176404,doi:10.1021/acs.nanolett.8b03321,PhysRevB.99.144401,PhysRevB.105.035109}. For 
multilayer graphene, it was shown that different (chiral) stacking creates rather distinct low-energy 
descriptions for the electron bands~\cite{RevModPhys.81.109,10.1143/PTPS.176.227,doi:10.1021/acs.nanolett.6b04698}, and thus leads to distinct and interesting electronic properties~\cite{Bao2011,Lui2011,PhysRevB.84.161408,doi:10.1126/sciadv.aat0074}. 
In the electronic systems, the stacking order changes the electronic properties 
by modifying the electron tunneling channels and the electron interactions. 
In magnets, the stacking order of the magnetic layers influences 
the lattice structure and then the magnetic interaction. 
Among many different possible stacking orders, we here choose an 
ABC stacking of the triangular layers. 
This choice turns out to be one 
of the simplest stackings that could generate magnetic frustration and non-trivial magnetic physics.
Clearly, the AA stacking is a simple uniform stacking along the $c$ direction and does not really lead to 
anything interesting if only the nearest-neighbor (NN) interaction is considered. 
The AB stacking, where the reference site of the B layer is projected to the center of the triangular 
plaquette on the A layer, generates interesting magnetic correlations and belongs to the extensively studied bipartite lattices. 
The ABC stacking in Fig.~\ref{fig:lattice_and_groundstates}(a), that seemingly triples the crystal unit cell, 
is in fact a 3D Bravais lattice. By creating a corner-shared tetrahedral structure along the $c$ axis,
the ABC stacking drastically enhances the magnetic frustration and can induce 
a classical spin-liquid regime at low temperatures even for Ising spins~\cite{PhysRevB.94.224413}. 
Together with the intralayer interaction 
from the ABC-stacked structure, the interlayer interactions 
generate rich and interesting magnetic behaviors including the subextensive ground-state degeneracy, 
thermal order-by-disorder, magnetic transition to spiral orders, 
thermal crossover and spiral spin liquid (SSL) regimes.  
We reveal these behaviors with the intralayer and interlayer 
Heisenberg interactions using a set of analytical techniques.

For each site of the ABC-stacking triangular multilayers, 
there exist six NN sites within the same layer and three in each of the two adjacent layers.
Distinct from the AB-stacking case, the triangular layer is no longer a mirror plane in the ABC-stacked case. 
Instead, the lattice site becomes an inversion center. 
The primitive lattice vectors are chosen as ${\bm{a}_1=(1,0,0)}$, ${\bm{a}_2=(-1/2,\sqrt{3}/2,0)}$, 
${\bm{a}_3=(1/2,\sqrt{3}/6,h)}$, where the interlayer separation $h$ varies for different materials. 
In this Letter, we take a unit layer distance ${h=1}$ for convenience. 
Starting from the NN antiferromagnetic Heisenberg model 
on the triangular lattice, we incorporate the NN interlayer spin interactions with the Hamiltonian
\begin{equation}\label{eq:hamiltonian}
	\mathcal{H} =  J_1\sum_{\braket{ij}_{\parallel} } \mathbf{S}_i \cdot \mathbf{S}_j + J_{\perp} \sum_{\braket{ ij}_\perp} \mathbf{S}_i \cdot \mathbf{S}_j. 
\end{equation}
Here $\braket{ij}_{\parallel}$ and $\braket{ij}_{\perp}$ refer to intra- and interlayer NN pairs, respectively. 
The antiferromagnetic interactions are denoted by $J_1$ and $J_{\perp}$ [see Fig.~\ref{fig:lattice_and_groundstates}(a)]. 
In the decoupling limit where ${J_{\perp}/J_1=0}$, the ground state on the monolayer triangular lattice is the well-known $120^{\circ}$ 
state. As we demonstrate below, the ABC stacking drastically enhances the magnetic frustration
and suppresses the magnetic ordering once the interlayer coupling is considered.

\emph{Zero-temperature classical ground states.}---By performing the Fourier transformation on the spin operator 
${\mathbf{S}_i=\frac{1}{\sqrt{N_s}}\sum_{\bm{k}}\mathbf{S}_{\bm{k}}\mathrm{e}^{\imath \bm{k} \cdot \bm{r}_i} }$, 
the spin Hamiltonian can be recast in 
the reciprocal space as ${\mathcal{H}= \sum_{\bm{k}}\mathbf{S}_{-\bm{k}}\mathcal{J}(\bm{k}) \mathbf{S}_{\bm{k}}}$, 
where $N_s$ is the total number of spins, ${\mathcal{J}(\bm{k})= \sum_{\bm{d}_{ij}}J_{ij}\mathrm{e}^{\imath\bm{k}\cdot \bm{d}_{ij}}}$ is the exchange interaction, and ${\bm{d}_{ij}\equiv \bm{r}_i-\bm{r}_j}$ denotes the NN vectors for both intra and interlayer bonds. 
Following the recipe of the Luttinger-Tisza method, this local unit-length constraint ${|\mathbf{S}_i|=1}$ for each spin is softened 
and replaced by a global one ${\sum_i |\mathbf{S}_i|=N_s}$. 
The classical ground state of the spin Hamiltonian can be obtained by searching the minimum eigenvalues 
of $\mathcal{J}(\bm{k})$ and verifying the satisfaction of the local constraints. 
It is convenient to introduce a complex parameter 
${\xi(\bm{k}) \equiv \Lambda(\bm{k})\mathrm{e}^{\imath \theta(\bm{k})} 
= 1+\mathrm{e}^{\imath \bm{k} \cdot \bm{a}_1} + \mathrm{e}^{\imath\bm{k} \cdot (\bm{a}_1 + \bm{a}_2)}}$, 
where its modulus and argument have been assigned to be $\Lambda(\bm{k})$ and $\theta(\bm{k})$, respectively. 
The exchange interaction is further rewritten as 
\begin{equation}
\label{eq:Jk}
	\mathcal{J}(\bm{k}) = \frac{1}{2}J_1[\Lambda(\bm{k})^2-3]+J_{\perp}\Lambda(\bm{k})\cos[\bm{k}\cdot\bm{a}_3-\theta(\bm{k})].
\end{equation}
At this stage, the minima of $\mathcal{J}(\bm{k})$ are simply characterized 
by ${ \xi(\bm{k})=- \mathrm{e}^{\imath\bm{k}\cdot \bm{a}_3} J_\perp/J_1 }$. 
By solving the equation about $\xi(\bm{k})$, the propagation vectors of the  
eigenvalue minima form several 1D manifolds in the reciprocal space 
for ${0 < J_{\perp}/ J_1 < 3}$ as shown in Figs.~\ref{fig:lattice_and_groundstates}(b-d). 
In particular, a spin-spiral state can be constructed through these propagation 
vectors and satisfies the local constraints strictly. 
Therefore, the spiral manifolds with a subextensive degeneracy from the Luttinger-Tisza 
method are the physical ground states. They are responsible for the formation of the SSL 
of the $(d_s,d_c)=(1,2)$ type~\cite{Bergman2007,Yao2021} at finite temperatures when 
thermal fluctuations are introduced. 
Here $d_s$ and $d_c$ refer to the dimension and codimension of spiral manifolds, respectively.

The degenerate spiral manifold evolves with $J_{\perp}/J_1$. 
In the weak interlayer coupling regime where ${J_{\perp}/J_1<1}$, 
the spiral manifolds manifest as six helices in Fig.~\ref{fig:lattice_and_groundstates}(b). 
Their projections onto the $k_x$-$k_y$ plane are comprised of six disconnected contours 
around the $K$ points in the Brillouin zone (BZ) for the monolayer triangular system.  
As $J_\perp/J_1$ increases from $0$ to $1$, the helices and their projected contours 
expands concurrently. For ${J_{\perp}/J_1=1}$, the spiral manifolds cross each other 
and become intersected lines in Fig.~\ref{fig:lattice_and_groundstates}(c). The degeneracy 
of the ground states reaches its maximum as well and indicates the strongest magnetic frustration. 
In the strong interlayer coupling regime with ${1 < J_{\perp}/J_1 < 3}$, the degenerate spiral manifold is 
further reduced into discrete and distort contours as shown in Fig.~\ref{fig:lattice_and_groundstates}(d). 
Their contours decrease with increasing $J_{\perp}/J_1$. Finally, they shrink into the points at $(0,0,\pm\pi)$ 
when ${J_{\perp}/J_1 \ge 3}$. The ground state turns out to be the antiferromagnetic (ferromagnetic) 
order between (within) the triangular layers.

\begin{figure*}
	\centering
	\includegraphics[width=1.0\textwidth]{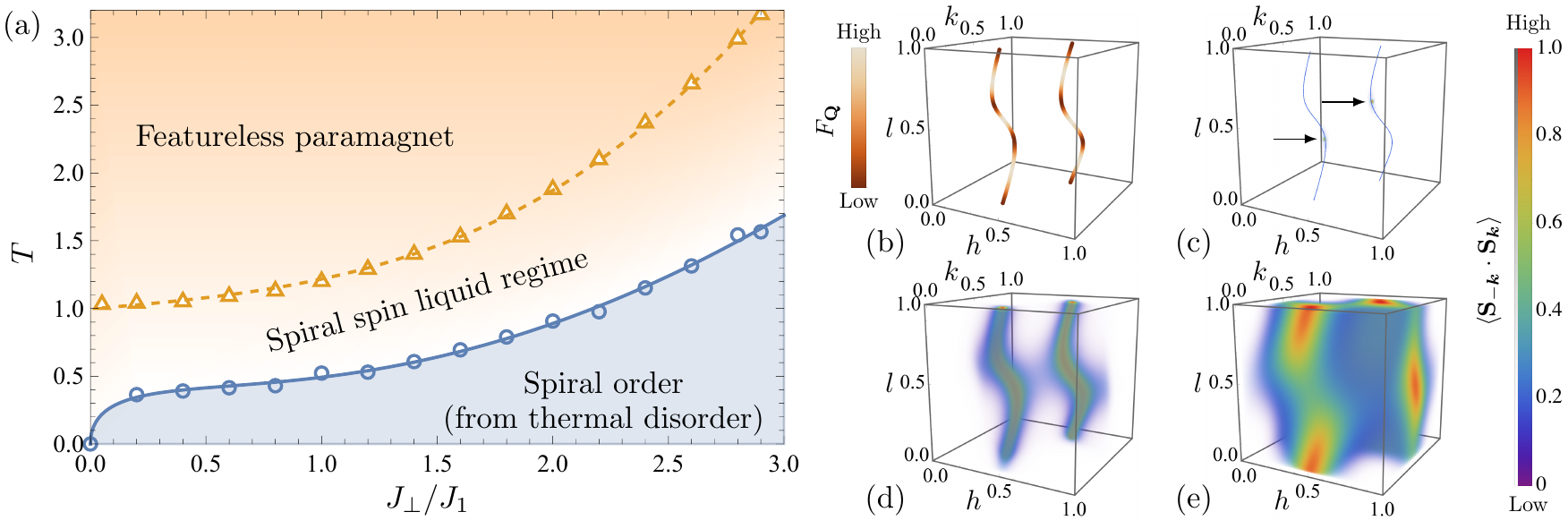}
	\caption{(a) The classical phase diagram for the $J_1$-$J_{\perp}$ Heisenberg model on an ABC-stacked triangular lattice. 
	The crossover (first-order phase transition) is outlined by the dashed (solid) line.
	(b) The distribution of free energy $F_{\mathbf{Q}}$ on the spiral manifolds for ${J_{\perp}/J_1=0.5}$. 
	The NBT results of ${\braket{\mathbf{S}_{\bm{-k}} \cdot \mathbf{S}_{\bm{k}}}}$ in (c) the spiral ordered phase with ${T=0.229}$, 
	(d) the SSL regime with ${T=0.429}$, and (e) the high-temperature paramagnet with ${T=1.589}$. 
	The regions with the lower density are set to be more transparent. 
	The arrows in (b) indicates the positions where $\braket{\mathbf{S}_{\bm{-k}} \cdot \mathbf{S}_{\bm{k}}}$ are highly concentrated on the spiral manifolds (blue).
	The system size is $50 \times 50 \times 50$.}%
	\label{fig:results_J1Jp}
\end{figure*}
 
\emph{Thermal order-by-disorder.}---As the temperature increases from absolute zero, the
thermal fluctuations enter into the system and could lift the subextensive ground-state degeneracy. 
For weak thermal fluctuations at low temperatures, this induces a discrepancy in the entropy 
for the spin-spiral wavevector on the the spiral manifold, despite the fact that 
different spin spiral configurations share the same energy. The one that possesses 
the highest entropy would be stabilized. This mechanism for the establishment of 
the long-range orders is known as the thermal 
order-by-disorder~\cite{Villain1980,PhysRevLett.62.2056,PhysRevB.48.9539,Bergman2007}. 
To formulate this effect for our case, we perform the low-temperature free energy and entropy
calculation, and the details can be found in the Supplemental Material (SM)~\cite{SM}. \nocite{Bergman2007,PhysRevLett.119.157202,PhysRevB.99.174404,PhysRevB.104.184427,PhysRevB.94.224413,PhysRevB.92.220417,PhysRevLett.93.167204,PhysRevResearch.4.013121,Yao2021,PhysRevB.40.7019,Gvozdikova2005}
In Fig.~\ref{fig:results_J1Jp}(a), we further depict the phase diagram 
and mark the regimes of thermal order by disorder. The finite temperature SSL 
regime is discussed in the later part of the Letter.

We sketch the thermal order-by-disorder effect here. 
At low temperatures, the thermal fluctuations of the spins are around the ground-state manifold. 
To characterize the thermal fluctuation of the spins, it is more convenient to parameterize the 
fluctuating spins based on the spin configurations from the ground-state manifold. 
For an arbitrary spin-spiral order with wavevector $\mathbf{Q}$, 
the spins would deviate from their ordered orientations 
${\bar{\mathbf{S}}_i=[\cos(\mathbf{Q} \cdot \bm{r}_i), \sin(\mathbf{Q} \cdot \bm{r}_i),0]}$ 
due to the thermal fluctuations. This deviation can be described by a perpendicular vector 
$\bm{\phi}_i$ as ${\mathbf{S}_i = \bm{\phi}_i+\bar{\mathbf{S}}_i(1-\phi_i^2)^{1/2} }$, 
and ${ |\phi_i| \ll 1 }$ at very low temperatures. 
To capture the low-temperature properties, it is sufficient to expand the Hamiltonian 
up to the quadratic order of the in-plane and out-of-plane components $\phi_i^{\text{i}}$ and 
$\phi_i^{\text{o}}$ with ${ \mathcal{H}_\phi=\sum_{ij} \tilde{J}_{ij}\phi_i^{\text{o}}\phi_j^{\text{o}}
+\tilde{J}_{ij}(\bar{\mathbf{S}}_i \cdot \bar{\mathbf{S}}_j)\phi_i^{\text{i}}\phi_j^{\text{i}} }$ 
and ${\tilde{J}_{ij} = J_{ij} - \delta_{ij}\mathcal{J}(\mathbf{Q})}$. Under this approximation,
the low-temperature free energy is given by 
\begin{equation}
	F_{\mathbf{Q}} \sim T\int_{\bm{k}} \ln W_{\mathbf{Q}}(\bm{k}) + C,
\end{equation}
where $W_{\mathbf{Q}}(\bm{k}) = - \mathcal{J}(\mathbf{Q}) 
+ \sum_{\bm{d}_{ij}} J_{ij}\mathrm{e}^{\imath \bm{k}   
\cdot \bm{d}_{ij}}\cos(\mathbf{Q} \cdot \bm{d}_{ij})$ 
and $C$ is a constant. 
In Fig.~\ref{fig:results_J1Jp}(b), we plot the distribution of $\mathbf{Q}$-dependent 
free energy $F_{\mathbf{Q}}$ on the spiral manifolds for ${J_{\perp}/J_1=0.5}$. 
The relative strength of $F_{\mathbf{Q}}$ is encoded into the color gradient,
and the darkest points represent the selected wave vectors
whose exact coordinates have been listed in the SM~\cite{SM}.

\emph{The finite-temperature behaviors.}---Upon further increasing the temperatures, the selected    
spin spiral orders via the thermal order-by-disorder would melt under the strong thermal fluctuations. 
Before entering into a featureless paramagnet, the SSL could be revived 
at intermediate temperatures. To fully reveal the finite-temperature behaviors, we here
implement a nematic bond theory (NBT)~\cite{PhysRevLett.119.157202} 
and the conventional self-consistent Gaussian approximation (SCGA) 
to construct the classical phase diagram, which has been shown in Fig.~\ref{fig:results_J1Jp}(a). 
Both methods start from the partition function in the form of an imaginary-time functional integral
\begin{equation}
\label{eq:partition_function}
	{\mathcal Z} = \int \mathcal{D}[\mathbf{S}]  \mathcal{D}[\chi] \, \mathrm{e}^{-\beta \mathcal{H}}
	\, \mathrm{e}^{-\imath \beta \sum_i \chi_i (|\mathbf{S}_i|^2-1)},
\end{equation}
where the Lagrange multiplier $\chi_i$ serves as an auxiliary field to 
impose the local constraint and $\beta$ is the inverse of temperature.

\begin{figure}[t]
	\centering
	\includegraphics[width=1.0\linewidth]{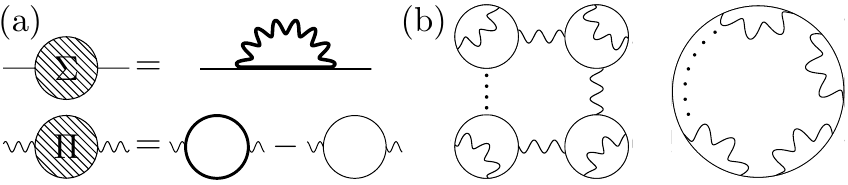}
	\caption{(a) Self-consistent equations for the self-energy $\Sigma$ and polarization $\Pi$. 
	(b) The derivative loop diagrams in the free energy density.}%
	\label{fig:feynman_diagram}
\end{figure}

In the NBT framework, the auxiliary constraint field $\chi_{\bm{k}-\bm{k}'}$ 
is divided into the static sector ${\Delta(T)=\imath\chi_{\bm{k}=0} }$ 
and the fluctuating sector ${ \Chi_{\bm{k},\bm{k}'} = - \imath \chi_{\bm{k}-\bm{k}'}(1-\delta_{\bm{k},\bm{k}'}) }$ 
after the Fourier transformation. The separation of variables yields the action 
\begin{equation}
\label{eq:full_action}
	\mathcal{S} = \beta \sum_{\bm{k},\bm{k'}} \mathbf{S}_{-\bm{k}}(K_{\bm{k},\bm{k}'} - \Chi_{\bm{k},\bm{k}'}) \cdot \mathbf{S}_{\bm{k}'} - \beta V \Delta(T),
\end{equation}
where ${K_{\bm{k},\bm{k}'} \equiv K_{0,\bm{k}}\delta_{\bm{k},\bm{k}'}=[\mathcal{J}(\bm{k})+\Delta(T)]\delta_{\bm{k},\bm{k}'} }$. 
An effective partition function ${\mathcal{Z} = \int \mathrm{d} \Delta \, \mathrm{e}^{\beta V \Delta(T)} \mathcal{Z}[\Delta] }$ 
can be obtained after the integration over the spin components in the large-$N$ limit~\cite{SM}. 
The effective action in $\mathcal{Z}[\Delta]$ is in the power of the field $\Chi$. 
To integrate the fluctuating sector $\Chi$ out, the self-consistent equations should 
be established for the bare spin propagators 
${ \braket{\mathbf{S}_{-\bm{k}}  \cdot \mathbf{S}_{\bm{k}}} 
 =  (2\beta)^{-1} NK_{0,\bm{k}}^{-1}}$ 
and the inverse constraint field propagators 
$\braket{\chi_{-\bm{k}} \, \chi_{\bm{k}}}^{-1} = D_{0,\bm{k}}^{-1} 
= N/2 \sum_{\bm{k}'} K_{0, \bm{k}+\bm{k}'}^{-1} K_{0, \bm{k}'}^{-1}$. 
They are renormalized perturbatively by the higher order $\Chi$ terms 
in $\mathcal{Z}[\Delta]$ and thus dressed by the a proper self-energy $\Sigma$ 
and polarization $\Pi$, respectively. The resulting Dyson equations are 
\begin{align}
	K_{\text{eff},\bm{k}} & = K_{0,\bm{k}} - \Sigma_{\bm{k}}, \\
	D^{-1}_{\text{eff},\bm{k}} & = D_{0,\bm{k}}^{-1} - \Pi_{\bm{k}}.  
\end{align}
As suggested in Ref.~\cite{PhysRevLett.119.157202}, at the cost of omitting all vertex corrections, 
the Dyson equations can be solved self-consistently  
with 
\begin{align}
	\Sigma_{\bm{k}} & = -\sum_{\bm{k}'\neq 0} K^{-1}_{\text{eff}, \bm{k}-\bm{k}'} D_{\text{eff},\bm{k}'},  \\ 
	\Pi_{\bm{k}} & = D_{0,\bm{k}}^{-1}-\frac{N}{2}\sum_{\bm{k}'} K^{-1}_{\text{eff},\bm{k}+\bm{k}'} K^{-1}_{\text{eff},\bm{k}'},
\end{align}
and are depicted as the diagrams in Fig.~\ref{fig:feynman_diagram}(a). 
With these approximations, the final integral in $\mathcal{Z}$ over the static sector $\Delta$ 
can be evaluated at the saddle point where ${NT/(2V)\sum_{\bm{k}}K_{\text{eff},\bm{k}}^{-1}=1}$. 
The free-energy density, that includes the loop diagrams in Fig.~\ref{fig:feynman_diagram}(b), 
are derived explicitly~\cite{SM}.

For the concerned parameter regime ${0 < J_{\perp}/J_1 < 3}$ where there
exists a subextensive degeneracy, the concentrated weights of spin structure 
factors $\braket{\mathbf{S}_{\bm{-k}} \cdot \mathbf{S}_{\bm{k}}}$, that are 
calculated with the NBT, are found in the low-temperature regime 
at the discrete momentum points, indicating the spin-spiral orders. 
As shown in Fig.~\ref{fig:results_J1Jp}(c), the positions of these 
high weights are identical to the results based 
on the entropy and the thermal order-by-disorder calculations. 
Moreover, the free energy density manifests a first-order phase transition 
above the ordered states at the temperatures shown in Fig.~\ref{fig:results_J1Jp}(a). 
The distribution of $\braket{\mathbf{S}_{\bm{-k}}  \cdot \mathbf{S}_{\bm{k}}}$ 
also changes drastically. Right above the transition temperature $T_C$, 
the point-like concentrations of $\braket{\mathbf{S}_{\bm{-k}} \cdot \mathbf{S}_{\bm{k}}}$ 
disappear immediately. Instead, there are clear spectral weight enhancements around 
the spiral manifold, and they decay rapidly away from it as shown in Fig.~\ref{fig:results_J1Jp}(d). 
These features are characteristic to the SSL~\cite{Bergman2007,Yao2021} 
and persist within a broad temperature window [see Fig.~\ref{fig:results_J1Jp}(a)].

The SSL behaviors are gradually overwhelmed with the prevailing thermal fluctuations. 
At higher temperatures, the spectral weights of ${\braket{\mathbf{S}_{\bm{-k}} \cdot \mathbf{S}_{\bm{k}}} }$ 
tend to spread throughout the whole BZ as shown in Fig.~\ref{fig:results_J1Jp}(e). 
Eventually, the spectral peaks around the spiral manifolds would become indiscernible 
when the system is deeply in the featureless paramagnet. The system experiences a crossover 
from the SSL to the featureless paramagnet. In the description of the NBT, 
the fluctuating sector $\Chi_{\bm{k},\bm{k}'}$ of the constraint field becomes insignificant 
and can be neglected in Eq.~\eqref{eq:full_action}. This simplification in the NBT leads to 
the well-known SCGA, which can qualitatively describe this thermal crossover~\cite{SM}. 
In the phase diagram of Fig.~\ref{fig:results_J1Jp}(a), the crossover temperatures 
are outlined based on the ``smoothening'' of the spectral peaks~\cite{SM}. Physically, 
this thermal crossover from higher temperatures to lower temperatures 
corresponds to the growth of the spin correlation. At a temperature  
much above Curie temperature, all the spins are fluctuating thermally
and there is not much correlation between the spins. At the order of the Curie temperature,
the spins become gradually correlated. At even lower temperatures in the SSL 
regime, the spin correlation in the the momentum space reveals the structures of the
degenerate spiral manifold. In the SSL regime, the thermal fluctuations are 
mainly around the spiral manifold, which may resemble the thermal fluctuation 
near a critical point to some extent, and a semi-universal thermodynamic 
property is expected. It is found that, the specific heat behaves like ${C_V= c_1 + c_2 T}$ 
in the SSL regime, where $c_{1,2}$ are constants~\cite{SM}.

\begin{figure*}[t]
	\includegraphics[width=1.0\linewidth]{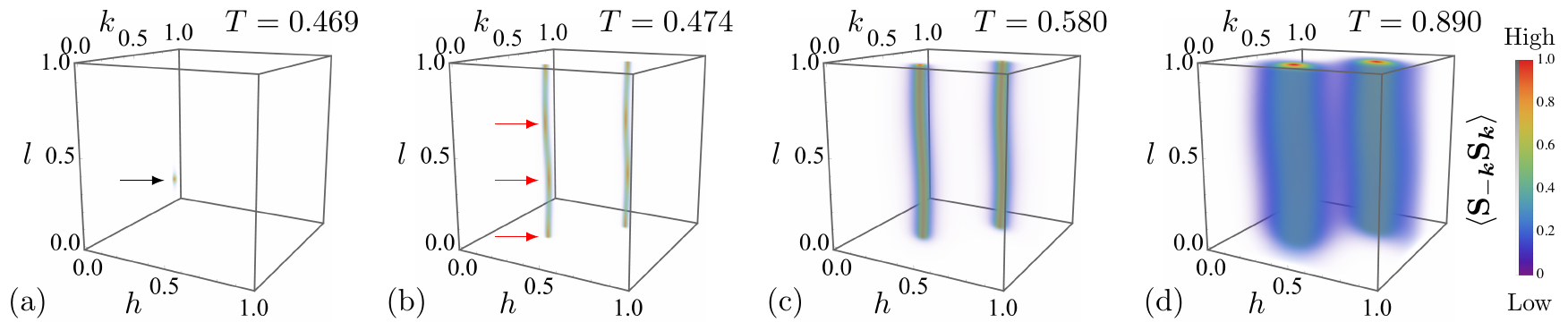}
	\caption{The NBT results of ${\braket{\mathbf{S}_{\bm{-k}} \cdot \mathbf{S}_{\bm{k}}} }$  
	at ${(J_1, J_\perp, J_2, J_3)=(1.0, 0.1, 0.0, 0.13)}$ for the system size ${50 \times 50 \times 50}$. 
	The first two temperatures are very close to the first-order transition temperature ${T_C \approx 0.470}$. 
	The arrows indicate the (a) point-like and (b) arc-shape concentrations of $\braket{\mathbf{S}_{\bm{-k}} \cdot \mathbf{S}_{\bm{k}}}$, respectively. 
	}%
	\label{fig:correlation_J1J2J3Jp}
\end{figure*}

\emph{Subleading spin interactions.}---While the thermal order-by-disorder and 
the entropy effect could lift the degeneracy of the spiral manifold at low temperatures,  
it is well-known that, other subleading spin interactions could enter and break the degeneracy.
For instance, in the presence of the second- and third-nearest spin interactions 
(denoted as $J_2$ and $J_3$, respectively), the spiral manifolds only exist 
at a special point ${J_2/J_1 = 2 J_3/J_1}$ and ${0 < J_{\perp}/J_1 \le 3 + 30 J_3/J_1}$~\cite{SM}. 
While this effect is clearly important at low temperatures, especially in the relevant 
\ce{ACrO2} antiferromagnets~\cite{PhysRevB.104.104422}, the more tempting question is about 
the stability of the SSL regime that is connected to the degenerate spiral manifold. 
Or, more experimentally, can the degenerate spiral manifold still manifest itself 
in the finite-temperature spin correlation? Certainly, when the subleading interaction
is rather weak, this is expected. To what extent the spin correlation is modified
by the subleading interaction, however, depends on the several competing energy 
scales and could vary from material to material.  
It is, therefore, more appropriate to simply demonstrate this for the specific interactions 
that are relevant to certain materials. 
We have performed the NBT calculations for ${(J_1, J_2, J_3, J_{\perp})=(1.0, 0.0, 0.13, 0.1)}$ 
that are closely relevant to the first-principles results for $\alpha$-\ce{HCrO2}~\cite{PhysRevB.104.104422}. 
The spin-spiral orders at low-temperatures are confirmed through the magnetic Bragg peak of 
$\braket{\mathbf{S}_{\bm{-k}} \cdot \mathbf{S}_{\bm{k}}}$ [see Fig.~\ref{fig:correlation_J1J2J3Jp}(a)]. 
A first-order transition is evidenced at ${T_C \approx 0.470}$~\cite{SM}. 

The spectral weights of $\braket{\mathbf{S}_{\bm{-k}} \cdot \mathbf{S}_{\bm{k}}}$ become 
pronounced along the degenerate spiral manifolds once the temperature exceeds $T_C$.
Its specific thermal evolution, however, carries a bit more structure. Within a narrow window 
${T_C < T \lesssim 0.573 }$, the most prominent weights appear near the ordered wave vectors 
[indicated by arrows in Fig.~\ref{fig:correlation_J1J2J3Jp}(b)]. 
With increasing temperature, two consecutive crossovers can be identified. 
First, the inhomogeneity of $\braket{\mathbf{S}_{\bm{-k}} \cdot \mathbf{S}_{\bm{k}}}$ 
along the degenerate spiral manifold is quickly flattened with the growing thermal fluctuations. 
A more homogeneous distribution is recovered when ${T \gtrsim 0.573}$, 
as shown in Fig.~\ref{fig:correlation_J1J2J3Jp}(c). Finally, the system undergoes 
another crossover into the featureless paramagnet, as indicated by the spreading 
of $\braket{\mathbf{S}_{\bm{-k}} \cdot \mathbf{S}_{\bm{k}}}$ in Fig.~\ref{fig:correlation_J1J2J3Jp}(d).

\emph{Discussion.}---The $J_1$-$J_{\perp}$ Heisenberg model for the SSL physics 
is quite distinct from previous studies based on bipartite lattices~\cite{Niggemann_2019,Yao2021}. 
Due to the geometric frustrations that are naturally induced by the ABC stacking,
an \textit{infinitesimal} interlayer coupling is sufficient to spawn the SSL. 
For bipartite lattice models, a finite interaction threshold is required for the SSL.  
For example, the criteria are ${J_2/ J_1 > 1/6}$ for the honeycomb lattice~\cite{PhysRevB.81.214419}, 
${J_2 / J_1 > 1/8}$ for the diamond lattice~\cite{Bergman2007}, and more strictly 
$J{_2 / J_1 = 2 J_3 / J_1 > 1/4}$ for the square lattice~\cite{Niggemann_2019}. 
This restriction may challenge the realization of SSLs because 
further exchange interactions can be relatively weak in real materials. 
The SSL condition ${0 < J_{\perp}/J_1 < 3}$ for our model 
is immediately realized once the stacking structure is fabricated to a sufficient number of layers.

Even for few layers, the SSL physics is still expected. 
When descending to a bilayer, our model is equivalent to a $J_1$-$J_2$ Heisenberg model on a honeycomb lattice. 
Furthermore, for even numbers of layers, the ABC-stacked triangular lattice can be viewed as a multilayer honeycomb lattice still with the ABC stacking despite a displacement of two sublattices along the $c$ direction. 
Very recently, a 2D SSL has been advocated by neutron scattering measurements in a vdW honeycomb magnet \ce{FeCl3} with the same stacking~\cite{PhysRevLett.128.227201}. 
It is also immune to intricate interlayer couplings. 
Although the interlayer spin exchanges are different here, a similar SSL is promising, e.g., through appropriate stacking controls. The nature of a few-layer version of our model is worthy of further study. 

Besides the stacking fabrication of vdW materials, ABC-stacked triangular multilayer
magnets actually exist in nature. There are a family of magnets with the formula $AMX_2$
where $A$ is a monovalent metal, $M$ is a trivalent metal such as the transition metal ion Cr~\cite{PhysRevLett.97.167203,Liu_2021,PhysRevB.88.180401,PhysRevB.104.104422,PhysRevMaterials.6.094013} or the rare-earth ion~\cite{Liu_2018,Bordelon2019,PhysRevX.11.021044,PhysRevB.101.224427}, 
and $X$ is a chalcogen, and the rhombohedral vdW compounds $MX_2$ such as \ce{NiBr2} and \ce{NiI2}~\cite{PhysRevB.84.060406,PhysRevB.87.014429,Tokura_2014,Mak2019,PhysRevMaterials.3.044001}.
Both families of magnets could experience extra magnetic anisotropies 
beyond the simple Heisenberg model. The simplest and common anisotropy for the transition metal ions
such as \ce{Cr^{3+}} and \ce{Ni^{2+}} ions is the single-ion spin anisotropy. In the presence of the easy-plane anisotropy, 
it is still possible to construct the spiral orders within the XY plane, and the SSL physics is still expected. 
With the easy-axis spin anisotropy, one cannot construct spiral orders with Ising spins and thus the ground-state 
configurations are completely different. The thermal fluctuations, however, 
could violate the Ising constraint and induce the SSL 
regime~\cite{PhysRevB.92.220417,PhysRevResearch.4.013121}.  
Besides the characteristics as shown in Figs.~\ref{fig:results_J1Jp}(c-e),
the spin structure factors could possess a reciprocal kagom\'{e}-like structure 
from the competition between frustration and spin stiffness~\cite{PhysRevResearch.4.013121}. 
The magnetic anisotropy for the rare-earth chalcogenides $AMX_2$ 
is mainly the exchange anisotropy from the strong spin-orbit coupling. 
Because of the short-range orbitals of the $4f$ electrons, the spin exchange 
is most likely to be dominated by the intralayer interactions,
and the SSL physics due to the interlayer coupling is probably 
less relevant over there. The mechanical control such as twisting, bending, and stacking is an 
uprising control knob of the physical properties of quantum materials. We hope our work
to stimulate some interest in the stacking control of quantum magnets and materials.

\begin{acknowledgments}
	We thank Chun-Jiong Huang for useful discussions.
	This work is supported by the National Science Foundation of China with Grant No.~92065203, 
	the Ministry of Science and Technology of China with Grants 
	No.~2021YFA1400300, by the Shanghai Municipal Science and Technology Major Project with 
	Grant No.~2019SHZDZX01, by NNSF of China with No.~12174067, and 
	by the Research Grants Council of Hong Kong with General 
	Research Fund Grant No.~17306520.
\end{acknowledgments}

\bibliography{SSL-TriangularABC.bib}

\end{document}


\title{Supplemental Material for ``Stacking-induced magnetic frustration and spiral spin liquid''}

\author{Jianqiao Liu}
\thanks{These authors contributed equally.}
\affiliation{State Key Laboratory of Surface Physics and
Department of Physics, Fudan University, Shanghai 200433, China}

\author{Xu-Ping Yao}
\thanks{These authors contributed equally.}
\affiliation{Department of Physics and HKU-UCAS Joint Institute 
for Theoretical and Computational Physics at Hong Kong, 
The University of Hong Kong, Hong Kong, China}

\author{Gang Chen}
\email{gangchen@hku.hk}
\affiliation{Department of Physics and HKU-UCAS Joint Institute 
for Theoretical and Computational Physics at Hong Kong, 
The University of Hong Kong, Hong Kong, China}
\affiliation{The University of Hong Kong Shenzhen Institute of Research and Innovation, Shenzhen 518057, China}



\maketitle

\section{Details about thermal order-by-disorder}%
\label{SM:TbOD}

In this section we present the detailed calculations of the thermal order-by-disorder, which is firstly proposed in Ref.~\cite{Bergman2007}. 
Starting from an arbitrary order with the wave vector $\mathbf{Q}$ among the spiral manifolds of SSLs, one can parameterize the corresponding spin-spiral configuration as
\begin{equation}
	\bar{\mathbf{S}}_i=[\cos(\mathbf{Q} \cdot \bm{r}_i), \sin(\mathbf{Q} \cdot \bm{r}_i),0]. 
\end{equation}
We have assumed that the spin vectors lie in the $x$-$y$ plane for simplicity. 
At low temperatures, the spins can slightly deviate from their ordered orientations, resulting in a physical configuration which can be approximated as
\begin{equation}
	\mathbf{S}_i =\bm{\phi}_i +\bar{\mathbf{S}}_i(1-\phi_i^2)^{1/2}. 
\end{equation}
A perpendicular field $\bm{\phi}_i$ with $|\bm{\phi}_i| \ll 1$ has been introduced here to capture the weak thermal fluctuation. 
The fluctuation field is constrained to be perpendicular to the assumed spin order $\bar{\mathbf{S}}_i$, i.e.,
\begin{equation}
	\bm{\phi}_i\cdot \bar{\mathbf{S}}_i=0, 
\end{equation}
to ensure that the unit-length of spins is satisfied locally. 
Given that the assumed spin order is restricted in the $x$-$y$ plane, the fluctuation field can be further decomposed into the in-plane sector $\mathbf{\phi}^{\text{i}}_i$ and the out-of-plane sector $\mathbf{\phi}^{\text{o}}_i$ as 
\begin{equation}
	\bm{\phi}_i=\phi^{\text{i}}_i(\hat{z} \times \bar{\mathbf{S}}_i) + \mathbf{\phi}^{\text{o}}_i\hat{z}. 
\end{equation}
To the second order of fluctuation variables, the spin Hamiltonian can be expanded into 
\begin{equation}
	\mathcal{H}_\phi=\sum_{ij} \left[ \tilde{J}_{ij}\phi_i^{\text{o}}\phi_j^{\text{o}}+\tilde{J}_{ij}(\bar{\mathbf{S}}_i \cdot \bar{\mathbf{S}}_j)\phi_i^{\text{i}}\phi_j^{\text{i}}  \right],
\end{equation}
where $\tilde{J}_{ij}=J_{ij}-\delta_{ij}\mathcal{J}(\mathbf{Q})$. 
The exchange interaction
\begin{widetext}
\begin{align}
	\mathcal{J}(\bm{k}) & = 
		\frac{J_1}{2}\left \{ \Lambda(\bm{k}) + \frac{J_{\perp}}{J_1} \cos[\bm{k}\cdot\bm{a}_3 - \theta(\bm{k})] \right \}^2 - \frac{J_{\perp}^2}{2J_1} \cos^2[\bm{k}\cdot\bm{a}_3-\theta(\bm{k})] - \frac{3}{2}J_1 \\
	 	& = \frac{1}{2}J_1[\Lambda(\bm{k})^2-3]+J_{\perp}\Lambda(\bm{k})\cos[\bm{k}\cdot\bm{a}_3-\theta(\bm{k})], \label{eq:Jk}
\end{align}
\end{widetext}
has been defined in the main text.
The lower order terms have been ignored safely. 
With this approximation, the partition function becomes
\begin{equation}
	\mathcal{Z} = \int \mathcal{D} [\phi^{\text{o}}] \mathcal{D} [\phi^{\text{i}}] \mathrm{e}^{-\beta \sum_{\bm{k}}\left[ \phi^{\text{o}}_{-\bm{k}}\tilde{J}(k)\phi^{\text{o}}_{\bm{k}} + \phi^{\text{i}}_{-\bm{k}}W_{\mathbf{Q}}(\bm{k})\phi^{\text{i}}_{\bm{k}} \right]},
\end{equation}
where $\tilde{J}(\bm{k})$ is the Fourier transform of $\tilde{J}_{ij}$ and
\begin{equation}
	W_{\mathbf{Q}}(\bm{k}) = - \mathcal{J}(\mathbf{Q}) + \sum_{\bm{d}_{ij}} J_{ij}\mathrm{e}^{\imath \bm{k} \cdot \bm{d}_{ij}}\cos(\mathbf{Q} \cdot \bm{d}_{ij}). 
\end{equation} 
The summation variable $\bm{d}_{ij} \equiv \bm{r}_i - \bm{r}_j$ denotes the nearest-neighbor vectors for both intra and interlayer bonds. 
The modified interaction $\tilde{J}(\bm{k})$ is independent on the wave vector $\mathbf{Q}$ defined on the spiral manifolds of SSLs. 
On the contrary, the function $W_{\mathbf{Q}}(\bm{k})$ varies with the spin-spiral wave vector $\mathbf{Q}$.
Therefore, the $\mathbf{Q}$-dependent free energy is given by
\begin{align}
	F_{\mathbf{Q}} & \sim T \int_{\bm{k}} \bigg \{ \ln \bigg[\frac{\tilde{J}(\bm{k})}{2\pi T}\bigg] + \ln \bigg[\frac{W_{\mathbf{Q}}(\bm{k})}{2\pi T}\bigg]  \bigg \} \notag \\
	& \sim T\int_{\bm{k}} \ln W_{\mathbf{Q}}(\bm{k}) + C, \label{eq:TObD_free_energy} 
\end{align}
where $C$ is constant which is only dependent on the temperature. 
In the Eq.~\eqref{eq:TObD_free_energy}, the temperature merely affects the absolute values of the free energy $F_{\mathbf{Q}}$ and plays no role on the relative intensity on the spiral manifolds. 
This approximation only applies to sufficiently weak fluctuations. 
Nevertheless, it qualitatively reveals the formation of long-range orders in the low-temperature regime and provides a good starting point for a more complete analysis in the next section. 

\begin{figure}[t]
	\centering
	\includegraphics[width=8.6cm]{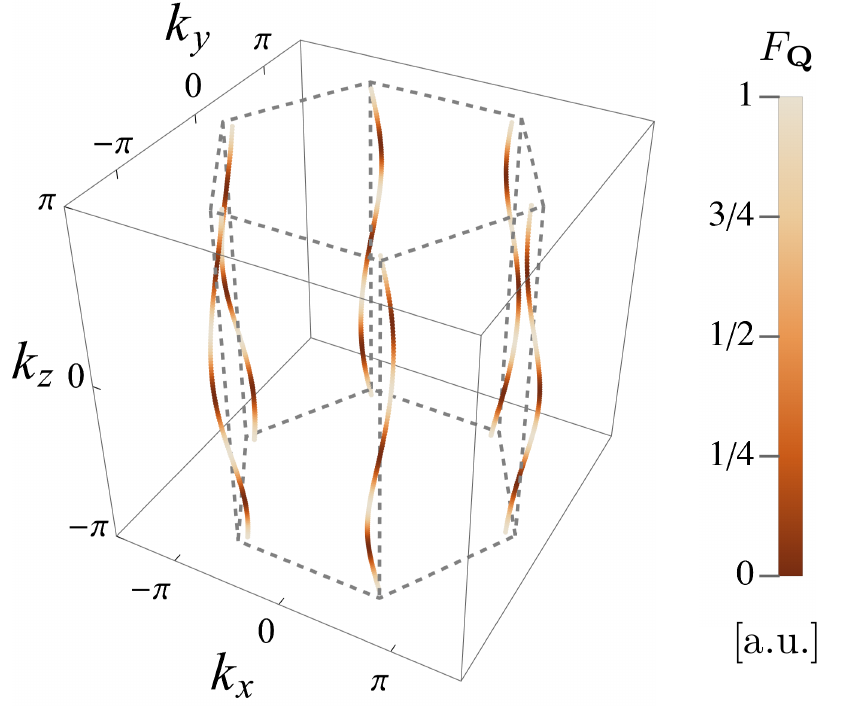}
	\caption{The distribution of $\mathbf{Q}$-dependent free energy $F_{\mathbf{Q}}$ for $J_\perp/J_1=0.25$. 
	The spiral manifolds are colored according to the relative strength of the free energy. 
	Dashed lines are sketched as a guide for the eye.
	In particular, the hexagon on the $k_x$-$k_y$ plane outlines the Brillouin-zone boundary of a monolayer triangular system.}%
	\label{fig:TObD}
\end{figure}

For the weak interlayer-coupling regime $0 < J_{\perp} / J_1 < 1$ as shown in Fig.~\ref{fig:TObD} and Fig.~3(a) of the main text, the minima of free energy $F_{\mathbf{Q}}$ locate at the following points in reciprocal space
\begin{align}
	\mathbf{Q}_1 & =\bigl( \pm 2 q, 2\pi / \sqrt{3}, 2\pi / 3 \bigr), \\
	\mathbf{Q}_2 & =\bigl( \pm q \mp \pi, \sqrt{3}q + \pi/\sqrt{3}, -2\pi/3 \bigr), \\
	\mathbf{Q}_3 & =\bigl( \pm q \mp \pi ,\sqrt{3}\pi - \sqrt{3}q,0 \bigr), 
\end{align}
where $q=\arccos\bigl(\frac{J_\perp}{2J_1}+\frac{1}{2}\bigr)$. 
For the strong interlayer coupling regime $1 \le J_{\perp}/J_1 < 3$, as shown in Figs.~\ref{fig:structure_factors} (a) and (e), the minima of free energy $F_{\mathbf{Q}}$ locate at 
\begin{align}
	\mathbf{Q}_1 & =\bigl( q_1, q_1 / \sqrt{3}, - q_2 - 2q_1 / 3 \bigr), \\
	\mathbf{Q}_2 & =\bigl( 0, 2q_1 / \sqrt{3}, - q_3 - q_1 / 3 \bigr), \\
	\mathbf{Q}_3 & =\bigl( q_1, - q_1 / \sqrt{3}, - q_3 - q_1 / 3 \bigr), 
\end{align}
where $q_1 = \arccos\bigl( \frac{J_{\perp}^2}{4J_1^2} - \frac{5}{4} \bigr)$, $q_2 = \arccos \bigl( \frac{3J_1}{2J_{\perp}} - \frac{J_{\perp}}{2J_1} \bigr)$, and $q_3 = \arccos \bigl( -\frac{3J_1}{4J_{\perp}} - \frac{J_{\perp}}{4J_1}\bigr)$. 
Other minima points can be obtained by applying the inversion, $C_3$ rotations, and lattice translations.

\section{Details about nematic bond theory}%
\label{SM:NBT}

The nematic bond theory (NBT) developed in Ref.~\cite{PhysRevLett.119.157202} is a large-$N$ expansion method where the large control parameter $N$ plays the role of the number of spin components. 
This generalization can significantly suppress the contribution of fluctuations near the saddle point, and hence render the integrals over the spin component Gaussian-type in the partition function. 
The realistic number $N = 3$ for classical Heisenberg spins will be retrieved after this procedure. 
This convention will be adopted implicitly hereafter unless otherwise specified. 
As an efficient self-consistent method for classical Heisenberg magnets, the NBT has been applied to both 2D and 3D lattices with large system sizes and obtained consistent results with classical Monte Carlo simulations~\cite{PhysRevB.99.174404,PhysRevB.104.184427}. 

\subsection{The NBT framework}

Specifically, the NBT starts by introducing a site-dependent auxiliary field $\chi_i$ to enforce the local constraint $|\mathbf{S}_i|^2=1$ for each spin variable. 
In the form of an imaginary-time path integral formulation, the corresponding partition function can be expressed as 
\begin{equation}\label{eq:Z_S_chi}
	\mathcal{Z} = \int \mathcal{D}[\mathbf{S}]  \mathcal{D}[\chi] \mathrm{e}^{-\beta \mathcal{H}}\mathrm{e}^{-\imath \beta \sum_i \chi_i (|\mathbf{S}_i|^2-1)}.
\end{equation}
The imposed constraint field $\chi_i$ serves as the Lagrange multiplier and $\beta$ is the inverse of temperature. 
By performing Fourier transformation on both spin $\mathbf{S}_i$ and constraint variable $\chi_i$, one can obtain the action in the exponent 
\begin{equation}
	\mathcal{S} =  \beta \sum_{\bm{k},\bm{k}'} \mathbf{S}_{-\bm{k}} [\mathcal{J}(\bm{k})\delta_{\bm{k},\bm{k}'} + \imath \chi_{\bm{k}-\bm{k}'}] \mathbf{S}_{\bm{k}'} - \imath \beta \chi_{\bm{k}=0}, 
\end{equation}
where the spin-exchange matrix $\mathcal{J}(\bm{k})$ has defined in Eq.~\eqref{eq:Jk}. 
The $\chi_{\bm{k}=0}$ term is uniform in real space and can be interpreted as the static sector of the constraint field. 
We introduce a new parameter $\Delta(T) = \imath \chi_{\bm{k}=0}$ to absorb the image coefficient and point out its temperature dependency explicitly. 
The fluctuating sector of the auxiliary constraint field, which is crucial to the fluctuations at intermediate temperatures, can be obtained immediately
\begin{equation}
	\Chi_{\bm{k},\bm{k}'} = -\imath \chi_{\bm{k}-\bm{k}'}(1 - \delta_{\bm{k},\bm{k}'}). 
\end{equation}
After this separation, the action in Eq.~\eqref{eq:Z_S_chi} can be recast into 
\begin{equation}\label{eq:action}
	\mathcal{S} =  \beta \sum_{\bm{k},\bm{k'}} \mathbf{S}_{-\bm{k}}(K_{\bm{k},\bm{k}'} - \Chi_{\bm{k},\bm{k}'})\mathbf{S}_{\bm{k}'} - \beta V \Delta(T),
\end{equation}
where
\begin{equation}
	K_{\bm{k},\bm{k}'} \equiv K_{0,\bm{k}}\delta_{\bm{k},\bm{k}'}=[\mathcal{J}(\bm{k})+\Delta(T)]\delta_{\bm{k},\bm{k}'},
\end{equation}
and $V$ is the volume of the spin system. 
Therefore, the constraint variables $\Chi_{\bm{k},\bm{k}'}$ can be viewed as a scalar auxiliary field coupling to the spin field by a Yukawa-like interaction. 
With this transformation, the integrals over the spin components in the partition function Eq.~\eqref{eq:Z_S_chi} are quadratic and can be finished in the large-$N$ limit. 
Formally, it yields
\begin{equation}
	\mathcal{Z} = \int \mathrm{d} \Delta \mathrm{e}^{\beta V \Delta(T)} \mathcal{Z}[\Delta],
\end{equation}
where $\mathcal{Z}[\Delta]$ is given by 
\begin{align}\label{eq:Z_Delta}
	\mathcal{Z}[\Delta] & = \int \mathcal{D} [\chi] \Bigl [\det(K-\Chi)^{-\frac{N}{2}}\Bigr ] \notag \\
	& = \int \mathcal{D} [\chi] \mathrm{e}^{-\frac{N}{2} \Tr \ln(K-\Chi)} \notag \\
	& = \int \mathcal{D} [\chi] \mathrm{e}^{-\frac{N}{2} \Tr \ln K +\frac{N}{2} \sum_{n=2}^\infty \frac{1}{n} \Tr(K^{-1}\Chi)^n}.
\end{align}
The effective action has been expanded in power of the fluctuating sector $\Chi$ of the constraint field because the fluctuations around the static constraint are assumed to be weak enough.

Now one can introduce the bare propagators for both spin and constraint fields as 
\begin{align}\label{eq:bare_propagators}
	\braket{\mathbf{S}_{-\bm{k}}\mathbf{S}_{\bm{k}}} & = \frac{N}{2\beta}K_{0, \bm{k}}^{-1}, \\
	\braket{\chi_{-\bm{k}}\chi_{\bm{k}}} & = D_{0,\bm{k}}. 
\end{align}
The latter can be derived from the effective action Eq.~\eqref{eq:Z_Delta} up to the second order in the field $\Chi$; it turns out to be the autocorrelation of $K^{-1}$
\begin{equation}\label{eq:constraint_propagator}
	D_{0,\bm{k}}^{-1} = \frac{N}{2}\sum_{\bm{k}'} K_{0, \bm{k}+\bm{k}'}^{-1} K_{0, \bm{k}'}^{-1}. 
\end{equation}
Higher order terms in the effective action Eq.~\eqref{eq:Z_Delta} represent the interaction effect in the theory. 
They would further renormalize the above propagators perturbatively, resulting in a proper self-energy $\Sigma$ and polarization $\Pi$ to the spin and constraint propagators, respectively. 
The dressed propagators can be expressed via the Dyson equations
\begin{align}
	K_{\text{eff},\bm{k}} & = K_{0,\bm{k}} - \Sigma_{\bm{k}}, \label{eq:Dyson_self_energy}\\
	D^{-1}_{\text{eff},\bm{k}} & = D_{0,\bm{k}}^{-1} - \Pi_{\bm{k}}. \label{eq:Dyson_polarization} 
\end{align}
The authors of Ref.~\cite{PhysRevLett.119.157202} suggested that at the cost of omitting all vertex corrections in the diagram, the Dyson equations can be solved self-consistently, inducing following elegant results
\begin{align}
	\Sigma_{\bm{k}} & = -\sum_{\bm{k}'\neq 0} K^{-1}_{\text{eff}, \bm{k}-\bm{k}'} D_{\text{eff},\bm{k}'}, \label{eq:self_energy} \\
	\Pi_{\bm{k}} & = D_{0,\bm{k}}^{-1}-\frac{N}{2}\sum_{\bm{k}'} K^{-1}_{\text{eff},\bm{k}+\bm{k}'} K^{-1}_{\text{eff},\bm{k}'}. \label{eq:polarization}
\end{align}
to the leading order of the parameter $N$. 
It can be seen that the self-energy is proportional to the convolution of $K_{\text{eff},\bm{k}}^{-1}$ and $D_{\text{eff},\bm{k}}$. 
Comparing the Eqs.~\eqref{eq:Dyson_polarization} and~\eqref{eq:polarization}, the dressed propagator $D_{\text{eff},\bm{k}}^{-1}$ of the constraint field holds the same autocorrelation form as the bare one in Eq.~\eqref{eq:constraint_propagator}.

With the approximated self-energy $\Sigma$ and poralization $\Pi$, the partition function can be reduced to 
\begin{equation}\label{eq:Z_delta}
	\mathcal{Z}_{\text{eff}} = \int \mathrm{d} \Delta \mathrm{e}^{-\mathcal{S}[\Delta]},
\end{equation}
with the effective action
\begin{equation}
	\mathcal{S}[\Delta] = \frac{N}{2} \Tr \ln K_{\text{eff}} + \frac{1}{2} \Tr \ln D_{\text{eff}}^{-1} + \frac{N}{2} \Tr (K_{\text{eff}}^{-1}\Sigma) - \beta V \Delta. 
\end{equation}
The final integral in $\mathcal{Z}_{\text{eff}}$ is taken over the static sector $\Delta$ of the constraint field and thus can be evaluated by the saddle point approximation. 
The saddle point condition is determined by differentiating $\mathcal{S}[\Delta]$, yielding  
\begin{equation}\label{eq:saddle_point}
	\frac{N T}{2V}\sum_{\bm{k}}K_{\text{eff},\bm{k}}^{-1}=1.
\end{equation}

\subsection{Free energy and specific heat}

The self-consistent equations from Eq.~\eqref{eq:bare_propagators} to Eq.~\eqref{eq:polarization}, together with the saddle point condition Eq.~\eqref{eq:saddle_point}, gives the explicit expression of free energy density $f=-1/(\beta V)\ln Z_{\text{eff}}$~\cite{PhysRevB.104.184427}
\begin{widetext}
\begin{equation}\label{eq:free_energy}
	f = -\Delta - \frac{N T}{2V}\sum_{\bm{k}}\ln(T K_{\text{eff}, \bm{k}}^{-1}) +\frac{NT}{2V}\sum_{\bm{k}}K_{\text{eff},\bm{k}}^{-1}\Sigma_{\bm{k}} + \frac{T}{2V}\sum_{\bm{k}} \ln\bigg[\frac{T^2 D_{\text{eff},\bm{k}}^{-1}}{2V}\bigg] - \frac{N-1}{2} T \ln \pi.
\end{equation}
\end{widetext}
The constant factors absorbed in the integration measures have been retrieved. 

The specific heat behavior can be analyzed by from the self-energy Eq.~\eqref{eq:self_energy} and the free energy Eq.~\eqref{eq:free_energy}.
Firstly, we recover the implicit factor $T$ in self-energy
\begin{equation}
	\Sigma_{\bm{k}} = -T\int_{\bm{k}^\prime \neq 0} \frac{D_{\text{eff},\bm{k}^\prime}}{K_{\text{eff},\bm{k}-\bm{k}^\prime}},
\end{equation}
and further assume its temperature dependence with the leading form $\Sigma_{\bm{k}}= T^\alpha \bar{\Sigma}_{\bm{k}}$. 
In the SSL regime, thermal fluctuations around the spiral manifold are dominant. 
These wave vectors constituting the spiral manifolds can be denoted as $\{\mathbf{Q}\}$. 
It is safe to replace the integration $\int_{\bm{k}^\prime}$ by $\int_{\bm{q} \in {\mathbf\{\mathbf{Q}\}}} \mathrm{d} \bm{q} \int \mathrm{d}^2\delta \bm{k}_n$. 
Here we have associated the wave vector $\bm{k}' \in \{\mathbf{Q}\}$ with a local coordination
system spanned by two normal vectors $\hat{\bm{n}}_{1,2}$ and one tangent vector $\hat{\bm{t}}$, where $\bm{k}'\cdot \hat{\bm{t}} = 0$ and $\delta\bm{k}_n$ in the $\hat{\bm{n}}_1$-$\hat{\bm{n}}_2$ plane.  
In the local coordinate system, any wave vector around spiral manifolds can be expressed as $\bm{k}' = \bm{q} +\delta \bm{k}_n$ and $|\delta\bm{k}_n|=\delta k$. 
With this transformation, the self-consistent equation reduces to 
\begin{equation}
	T^\alpha \bar{\Sigma}_{\bm{k}} = - T\int_{\bm{q} \in \{\mathbf{Q}\}} \int \mathrm{d}^2\delta \bm{k}_n \frac{D_{\text{eff},\bm{k}^\prime}}{v^2 (\delta k)^2 - T^\alpha \bar{\Sigma}_{\bm{k}}}. 
\end{equation}
The dispersion of the energy has been approximately expressed in a quadratic form $v^2 (\delta k)^2$ near the spiral manifolds. 
The integration over the $\hat{\bm{n}}_1$-$\hat{\bm{n}}_2$ plane should be implemented with a proper cutoff. 
By replacing the integral variable $\delta k = T^{\alpha/2} \delta p$, the temperature dependence can be scaled out with $\alpha = 1$. 
Given the relationship $\Sigma_{\bm{k}} = T \bar{\Sigma}_{\bm{k}}$, the temperature scaling of free energy density can be obtained
\begin{align}
	f & = -\Delta - \frac{N}{2} T \ln T + \frac{NT}{2V} \int_{\bm{k}} \ln [v^2 (\delta k)^2 + \Sigma_{\bm{k}}] + \frac{NT^2}{2V}\int_{\bm{k}} \frac{\bar{\Sigma}_{\bm{k}}}{v^2 (\delta k)^2 + T \bar{\Sigma}} + \frac{1}{2}T \ln T^2 \notag \\
	& \qquad \quad + \frac{T}{2V}\int_{\bm{k}} \ln \frac{D_{\text{eff},\bm{k}}^{-1}}{2V} - \frac{N-1}{2}T \ln \pi \notag \\
	& = c_1 T \ln T +c_2 T + c_3 T^2 + c_4 T \ln T^2. 
\end{align}
Therefore the specific heat in the SSL regime reads 
\begin{equation}
C_V = -T \frac{\partial^2 F}{\partial T^2} = (c_1+2c_4) + 2c_3 T,
\end{equation}
where all coefficients have been absorbed into constants $c_i$. 

\subsection{The numerical results}

Thanks to the efficiency of the NBT method, a rather large system size up to $50 \times 50 \times 50$ can be achieved in the self-consistent calculation. 
The temperature $T$ is monitored during the iteration procedure for the convergence. 
The convergence is reached when the criterion $|T_{i} - T_{i-1}| / T_i < 10^{-8}$ is met at the $i$th step. 
Within each iteration procedure, the parameters $\Delta$ take one of the values from $10^{-12}$ to $1$ evenly divided on a logarithmic scale. 
The implement of the self-consistent iteration also requires an initialization of the self energy $\Sigma_{\bm{k}}$, or equivalently the dressed spin propagator $K_{\text{eff},\bm{k}}$. 
In practice, We have adopted two initialisation schemes. 
The first one is to randomly initialize $\Sigma_{\bm{k}}$ and the second is to specify the values of $K_{\text{eff},\bm{k}}$ directly according to the predictions based on the thermal order-by-disorder calculations. 
Both two schemes give the almost identical results within the error due to the finite discretization of the Brillouin zone, if the convergence is reached.

\begin{figure}[t]
	\includegraphics[width=8.6cm]{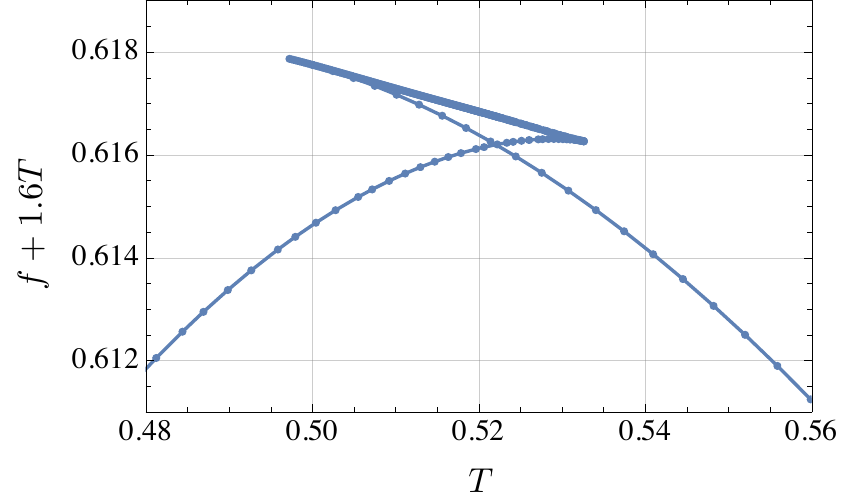}
	\caption{The evolution of free energy density $f$ (rescaled by adding a linear-$T$ term) with the temperature $T$ at $J_{\perp}/J_1 = 1.00$ for the system size $50 \times 50 \times 50$. 
	The free energy density is multivalued in a narrow regime near $T=0.52$. 
	The lowest branch should be accepted as the true results for the ground states, indicating a first-order phase transition at its cusp. 
	}%
	\label{fig:free_energy_density}
\end{figure}

\begin{figure*}[t]
	\centering
	\includegraphics[width=\linewidth]{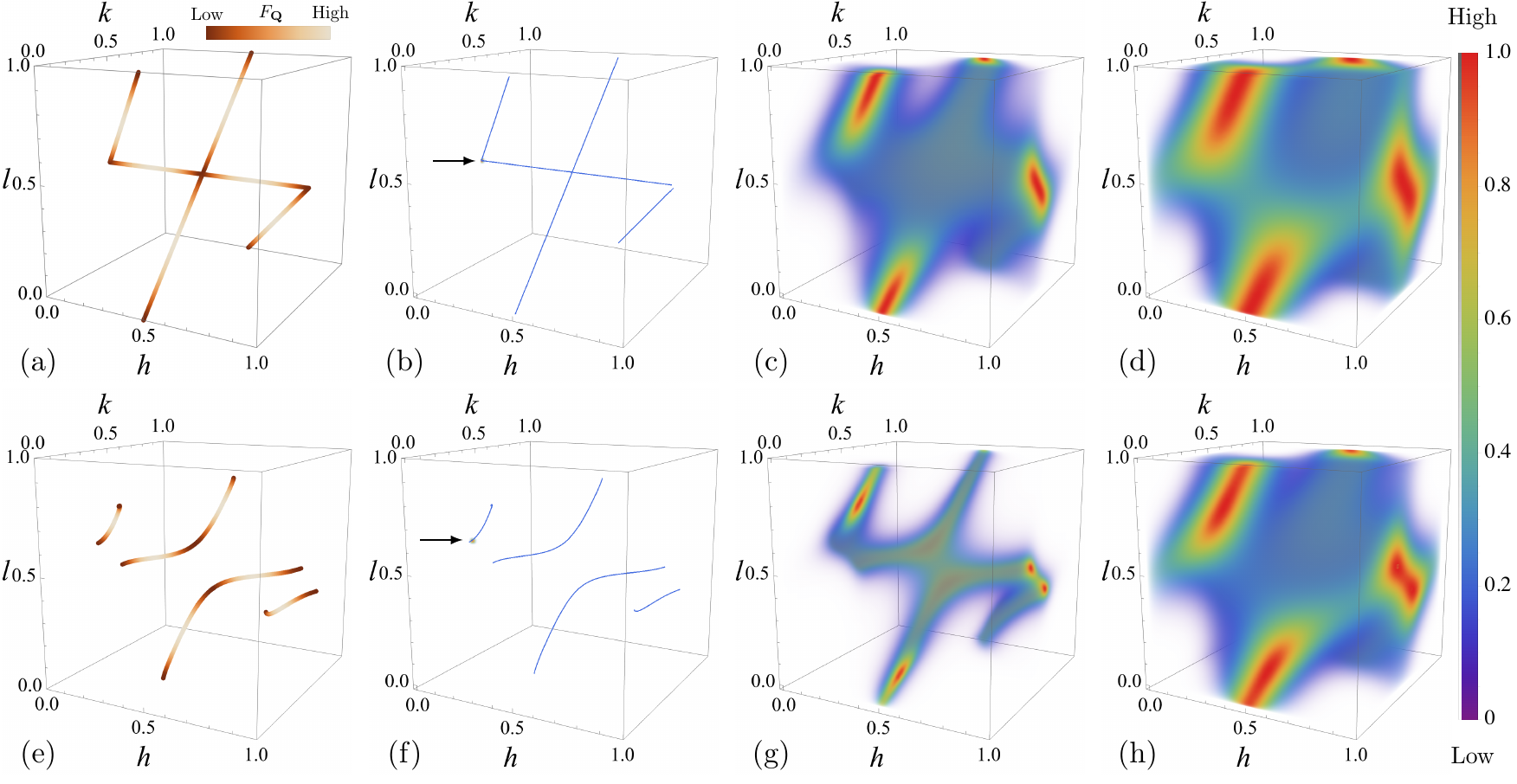}
	\caption{The distribution of free energy $F_{\mathbf{Q}}$ on the spiral manifolds for (a) $J_{\perp}/ J_1 = 0.5$ and (e) $J_{\perp}/J_1 = 1.0$. 
	The wave vectors with the lowest values are selected by the thermal order-by-disorder effect. 
	The second, third, and fourth columns are the NBT results of spectral weights $\braket{\mathbf{S}_{-\bm{k}}\mathbf{S}_{\bm{k}}}$ for the system size $50 \times 50 \times 50$ in the low-temperature ordered phase, the SSL regime, and the high-temperature paramagnets, respectively. 
	The corresponding temperatures are (b) $T=0.10$, (c) $T=0.54$, (d) $T=1.42$, (f) $T=0.20$, (g) $T=0.57$, and (h) $T=1.54$. 
	The regions with the lower density are set to be more transparent. 
	The arrows in (b) and (f) indicate the positions where spin structure factors are highly concentrated on the spiral manifolds (blue).
	}%
	\label{fig:structure_factors}
\end{figure*}

In Fig.~\ref{fig:free_energy_density}, we present the typical evolution of the free energy density as increasing temperature for the antiferromagnetic $J_1$-$J_{\perp}$ Heisenberg model at $J_1/J_{\perp} = 1.00$. 
For clarity, we have rescaled $f$ by adding linear-$T$ term to highlight its singular behavior near the phase transition point. 
In a narrow rage $0.495 \le T \le 0.535$, the free energy density is multivalued, which means there are more than one parameter $\Delta(T)$ satisfying the self-consistent equations. 
Physically, only the lowest branch is the thermodynamical stable solutions. 
The cusp identified in the free energy density reflects the discontinuity of its first derivative with respect to the temperature. 
Thus, it signals a first-order phase transition with jumped order parameters in the intermediate temperature regime.

The spin structure factors for different temperature regimes have been shown in Fig.~3 of the main text for $J_{\perp}/J_1=0.5$ and Fig.~\ref{fig:structure_factors} for $J_{\perp}/J_1 = 1.0$ and $1.3$. 
They represent three different forms of the degenerate spiral manifolds, respectively. 
The transition temperatures $T_c$ for different interactions $J_{\perp}/J_1$ have been determined in the same way and presented in the phase diagram Fig.~2 of the main text. 
The phase on the lower-temperature side should be nematic ordered in the NBT context, i.e., it breaks lattice symmetries spontaneously. 
Note that, in Figs.~\ref{fig:structure_factors} (b) and (f),the wave vectors of spiral ordered states are incommensurate and should appear in pairs. 
However, only one of them is clearly visible because of the finite discretization ($50 \times 50 \times 50$) of the Brillouin zone in the natural reciprocal space coordinates $(h, k, l)$. 
As we discussed in the main text, they are identical to the spiral wave vectors $\mathbf{Q}$ selected by the thermal order-by-disorder effect. 
In the nearby phase slightly above the first-order transition temperature $T_c$, the significantly enhanced spectral weights around the spiral manifolds are characteristic of SSLs.

\section{Details about self-consistent Gaussian approximation}%
\label{SM:SCGA}

At high temperatures, the fluctuations of the constraint field $\chi$ around its static values $\Delta(T)$ seem to be insignificant because of the strong thermal average. 
It is expected that all the lattice symmetries are preserved in the homogeneous paramagnetic state. 
Therefore, it is justifiable to neglect the matrix $\Chi_{\bm{k},\bm{k}'}$ in the action Eq.~\eqref{eq:action} because the fluctuating sector of the constraint field $\chi$ is responsible for the spontaneous breaking of lattice symmetries. 
The simplified partition function reads
\begin{equation}
	\mathcal{Z} = \int \mathcal{D}[\mathbf{S}] \mathcal{D}[\Delta] \mathrm{e}^{-\mathcal{S}_{\text{eff}}}, 
\end{equation}
where the effective action is reduced to 
\begin{equation}
	\mathcal{S}_{\text{eff}} = \beta \sum_{\bm{k}} \mathbf{S}_{-\bm{k}}[\mathcal{J}(\bm{k}) + \Delta(T)] \mathbf{S}_{\bm{k}} - \beta V \Delta(T). 
\end{equation}
This simplified method is the well-known self-consistent Gaussian approximation (SCGA). 
The SCGA has been widely used for classical frustrated magnets~\cite{PhysRevB.94.224413,PhysRevB.92.220417,PhysRevLett.93.167204} and is believed to serve as a qualitatively reasonable tool for these problems, especially at high temperatures~\cite{PhysRevResearch.4.013121}. 
Treating the uniform field $\Delta(T)$ as the saddle point parameter, the spin correlation correlation is still given by Eq.~\eqref{eq:bare_propagators} as 
\begin{equation}
	\braket{\mathbf{S}_{-\bm{k}}\mathbf{S}_{\bm{k}}} = \frac{N}{2\beta}\frac{1}{\mathcal{J}(\bm{k})+\Delta(T)}. 
\end{equation}
The self-consistent determination of the saddle point is more easer than that in Eq.~\eqref{eq:saddle_point},
\begin{equation}
	\frac{NT}{2V}\sum_{\bm{k}}\frac{1}{\mathcal{J}(\bm{k})+\Delta(T)} = 1. 
\end{equation}

\begin{figure}[t]
	\centering
	\includegraphics[width=8.6cm]{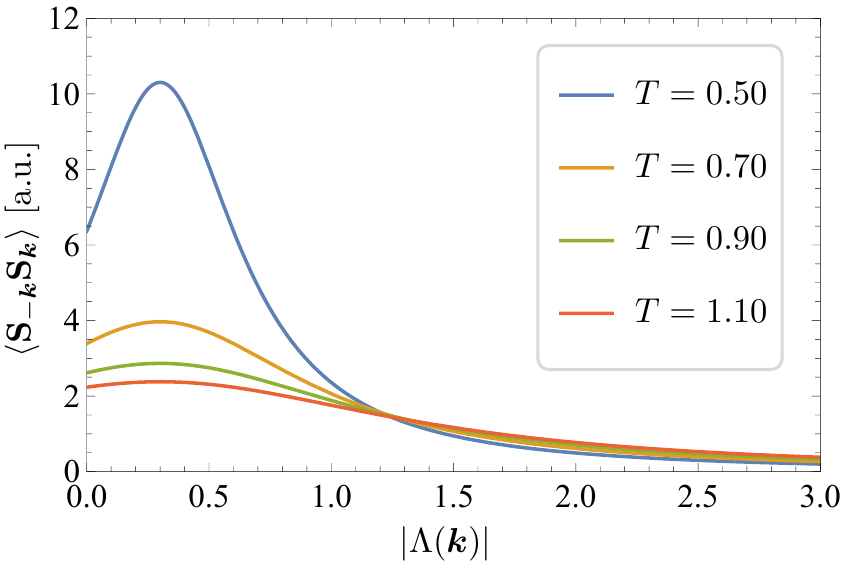}%
	\caption{Spin correlation $\braket{\mathbf{S}_{-\bm{k}}\mathbf{S}_{\bm{k}}}$ (where $\mu=x,y,z$) as a function of $|\Lambda(k)|$ at different temperatures. 
	The exchange interactions are set to be $J_{\perp}/J_1 = 0.30$, resulting in a spin correlation peak at $|\Lambda(\bm{k})| = 0.30$.}%
	\label{fig:crossover}
\end{figure}

As demonstrated in Refs.~\cite{Bergman2007,Yao2021,PhysRevResearch.4.013121}, the relative spectrum weights of the spin correlation function can be evaluated in the SCGA framework. 
We calculate the temperature evolution of the spin correlation $\braket{\mathbf{S}_{-\bm{k}}\mathbf{S}_{\bm{k}}}$ as a function of the parameter $|\Lambda(\bm{k})|$, where $\Lambda(\bm{k})$ is the modulus of $\xi(\bm{k})$; that is 
\begin{equation}\label{eq:xi}
	\xi(\bm{k}) \equiv \Lambda(\bm{k})\mathrm{e}^{\imath \theta(\bm{k})} = 1+\mathrm{e}^{\imath \bm{k} \cdot \bm{a}_1} + \mathrm{e}^{\imath\bm{k} \cdot (\bm{a}_1 + \bm{a}_2)}
\end{equation}
as defined in the main text. 
The results are depicted in Fig.~\ref{fig:crossover}. 
Because $\xi(\bm{k}) = -J_{\perp}/J_1 \mathrm{e}^{\imath\bm{k}\cdot\bm{a}_3}$, there is a peak of $\braket{\mathbf{S}_{-\bm{k}}\mathbf{S}_{\bm{k}}}$ near $|\Lambda(\bm{k})| = J_{\perp}/J_1$, which is obvious at the low temperature $T=0.50$. 
With the increasing of temperature, the correlation peaks near the spiral manifolds would be indiscernible gradually. 
The weights of the spin correlation $\braket{\mathbf{S}_{-\bm{k}}\mathbf{S}_{\bm{k}}}$ tends to spread throughout the whole Brillouin zone, indicating a crossover from the SSL to a featureless paramagnetic state. 
The estimated crossover temperatures for different interactions $J_{\perp}/J_1$ have been presented in the phase diagram Fig.~2 of the main text.

\section{Sprial spin liquids in the presence of further spin-exchange interactions}

For the ABC-stacked multilayer triangular magnets, the extensive ground-state degeneracy cannot exist without the fine-tuning when further spin-exchange interactions are incorporated. 
Here we further consider the second and third nearest spin-exchange interactions on top of the $J_1$-$J_{\perp}$ Heisenberg model. 
They are both intralayer Heisenberg interactions and can be denoted as $J_2$ and $J_3$, respectively. 
It is quite convenient to introduce new parameters 
\begin{align}
	A (\bm{k}) & = \cos(\bm{k}\cdot\bm{a}_1) + \cos(\bm{k}\cdot\bm{a}_2) + \cos[\bm{k}\cdot(\bm{a}_1+\bm{a}_2)] = \frac{1}{2}[\Lambda(\bm{k})^2 - 3], \\ 
	B (\bm{k}) & = \cos[\bm{k}\cdot(\bm{a}_1-\bm{a}_2)] + \cos[\bm{k}\cdot(\bm{a}_1+2\bm{a}_2)] + \cos[\bm{k}\cdot(2\bm{a}_1+\bm{a}_2)],
\end{align}
where $\Lambda(\bm{k})$ and $\theta(\bm{k})$ have been defined in main text and also in Eq.~\eqref{eq:xi}. 
Then the exchange matrix $\mathcal{J}(\bm{k})$ is generically given by 
\begin{multline}
	\mathcal{J}(\bm{k}) = 2J_3\left[ A(\bm{k}) - \frac{1}{2} \left( 1 - \frac{J_1}{2J_3} \right) \right]^2 + (J_2 - 2 J_3) B(\bm{k}) + J_{\perp} \Lambda(\bm{k}) \cos[\bm{k}\cdot\bm{a}_3 - \theta(\bm{k})] \\
	- \frac{J_3}{2}\left( 1 - \frac{J_1}{2 J_3}\right)^2 - 3J_3.
\end{multline}
At a special point $J_2 = 2 J_3$, the $B(\bm{k})$ term vanishes and the the spiral manifolds are restored similar to the analysis in the main text. 
The exchange matrix can be recast into
\begin{equation}
	\mathcal{J}(\bm{k}) = \frac{J_3}{2} \left[\Lambda^2(\bm{k}) + \frac{J_1}{2J_3} - 4 \right]^2 + J_\perp\Lambda(\bm{k})\cos\left[\bm{k}\cdot \bm{a}_3-\theta(\bm{k})\right],
\end{equation}
where a $\bm{k}$-independent constant has been dropped.

When $J_\perp=0$, the system degrades to decoupled layers and the wave vectors corresponding to the minimums of $\mathcal{J}(\bm{k})$ form contours in $k_x$-$k_y$ plane as long as $J_3 / J_1 > 1/8$. 
In the presence of finite $J_{\perp} >0$, the critical value of $J_3 / J_1$ becomes zero and the extensive degeneracy emerge when $0 < J_{\perp}/J_1 \le 3 + 30 J_3/J_1$. 
With the increasing of $J_{\perp}$, the spiral manifolds evolve from helices to contours similar to Figs~1(b-e) of the main text. 
Based on these continuous degenerated spiral orders, we expect that the SSL physics can emerge at intermediate temperatures for (or even near) the parameter region $J_2 = 2 J_3$.
The full phase diagram of the $J_1$-$J_2$-$J_3$-$J_\perp$ Heisenberg model in a wide parameter space are beyond the scope of this work and remained to be studied.

\begin{figure}[t]
	\includegraphics[width=8.6cm]{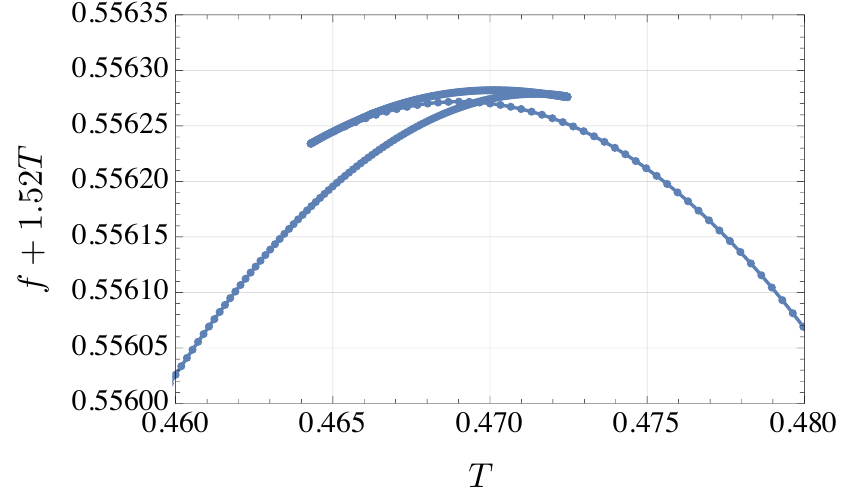}
	\caption{The evolution of free energy density $f$ (rescaled by adding a linear-$T$ term) 
	with the temperature $T$ at $(J_1, J_\perp, J_2, J_3)=(1.0, 0.1, 0, 0.13)$ 
	for the system size $50 \times 50 \times 50$.
	}%
	\label{fig:free_energy_density_J3}
\end{figure}

After incorporating the further spin-exchange interactions on the triangular layer, the SSL physics could persist at finite temperatures when the energy gap is not large enough comparing with the thermal fluctuations. 
In the presence of the second and third nearest-neighbor interactions $J_{2,3}$, we perform the NBT calculations for the parameters $(J_1,J_2,J_3,J_{\perp}) = (1.0, 0.0, 0.13, 0.1)$. 
The evolution of the free energy density is depicted in Fig.~\ref{fig:free_energy_density_J3}. 
It turns out that there is also a multivalued region between $0.464 \le T \le 0.473$, indicating a first-order phase transition at $T_c \approx 0.470$. 
The structure factors which possess more complicated features have been presented in Fig.~4 of the main text. 

\section{Connection with the FCC lattice antiferromagnet}

In general, the ABC-stacked multilayer triangular lattice is a 3D Bravais lattice belonging to the triclinic crystal system. 
For the primitive lattice vectors 
\begin{align}
	\bm{a}_1 & = (1,0,0), \\
	\bm{a}_2 & = (-1/2,\sqrt{3}/2,0), \\
	\bm{a}_3 & = (1/2,\sqrt{3}/6,h),
\end{align}
chosen in the main text, the reciprocal lattice vectors are 
\begin{align}
	\bm{b}_1 & = 2\pi (1,1/\sqrt{3}, -2/3), \\
	\bm{b}_2 & = 2\pi (0, 2/\sqrt{3}, -1/3), \\
	\bm{b}_3 & = 2\pi (0, 0, 1/h). 
\end{align}
There exist a special point where the interlayer distance becomes $h=\sqrt{2}/\sqrt{3}$. 
All three primitive lattice vectors have the unit length and the lattice symmetry is highly augmented. 
In fact, the ABC-stacked triangular lattice can be mapped to the face-centered-cubic (FCC) lattice immediately, which has been studied extensively. 
Here we clarify the connection between our results to the FCC antiferromagnet. 

For $J_{\perp}/J_1=1$, the $J_1$-$J_{\perp}$ Heisenberg model is identical to the nearest-neighbor antiferromagnetic Heisenberg model on the FCC lattice. 
As shown in Fig.~(c) of the main text, the spiral manifolds cross each other and become intersected lines in reciprocal space at $J_{\perp}/J_1=1$. 
The degeneracy of ground states reaches its maximum as well since the spiral manifold has the largest length. 
The most extensive degeneracy also indicates that the system has been promoted to a fully 3D one with the strongest magnetic frustration. 
This feature is consistent with the geometric structure of the FCC lattice. 
The transition temperature between the low-temperature order state and the SSL is about $T_c \approx 0.52$ from our NBT calculations. 
It is slightly higher than $T_c \approx 0.44$ from previous Monte Carlo simulations~\cite{PhysRevB.40.7019,Gvozdikova2005}. 
The discrepancy may be attributed to the approximations used in the NBT framework; that is only finite types of loop diagrams are included in the self-consistent calculations. 
It is expected to be reconciled after including the vertex corrections in the Feynman diagrams. 

When $J_{\perp}/J_1 \ge 3$, the spiral manifolds shrink into points at $(0,0,\pm\pi)$ as shown in Fig.~(e) of the main text. 
The ground state turns to be antiferromagnetic between adjacent layers and ferromagnetic within layers.
This result is reminiscent of the type-II antiferromagnetic order on the FCC lattice.

\bibliography{SSL-TriangularABC.bib}